\documentclass[a4paper,11pt]{article}
\usepackage{pos}

\title{Resummation effects in weak SUSY processes}

\author*[a]{Juri Fiaschi}
\author[a]{Michael Klasen}

\affiliation[a]{Institut für Theoretische Physik, Westfälische Wilhelms-Universität Münster, Wilhelm-Klemm-Straße 9, 48149 Münster, Germany}

\emailAdd{fiaschi@uni-muenster.de}
\emailAdd{michael.klasen@uni-muenster.de}

\abstract{
We present updated results for the production cross sections of slepton pairs and neutralino-chargino pairs at the LHC with next-to-next-to logarithmic precision matched at approximate QCD next-to-next-to leading order.
The explored range of masses of the supersymmetric particles are chosen to be relevant for current and future searches at the LHC.
We find moderate increases in the invariant mass distributions and integrated cross sections, and substantial reductions in the scale uncertainty of the results.
}

\FullConference{%
  40th International Conference on High Energy physics - ICHEP2020\\
  July 28 - August 6, 2020\\
  Prague, Czech Republic (virtual meeting)
}

\begin{document}

\begin{flushright}
MS-TP-20-38\\
\end{flushright}

\maketitle

\section{Introduction}
\vspace{-1em}

The Minimal Supersymmetric (SUSY) Standard Model (MSSM) is a very well motivated framework for physics beyond the Standard Model, since it offers elegant solutions of theoretical open questions, such as the stabilization of the Higgs mass and the unification of gauge couplings.
Furthermore, the lightest supersymmetric particle (LSP), usually the lightest neutralino, represents a natural dark matter candidate leading to a relic density consistent with the experimental observations.

SUSY searches at the LHC are constantly raising their sensitivity, as the various experiments, in particular ATLAS and CMS, continue to acquire data; for a consistent comparison with the theoretical predictions, it is then necessary that perturbative QCD calculations increase their precision by moving forward from leading order (LO) to next-to-leading order (NLO) and beyond.
In some kinematical regions, namely when heavy particles are produced close to threshold or with small transverse momentum, the logarithmic terms that appear in the expansion can become sizeable and require a resummation to all orders.

In this work, we update the predictions of the cross sections for sleptons pair~\cite{Fiaschi:2019zgh} and neutralino-chargino production~\cite{Fiaschi:2020udf} from next-to-leading logarithms (NLL) to next-to-next-to-leading logarithms (NNLL) precision matched at approximate NNLO QCD calculations (aNNLO), the latter containing only NLO SUSY-QCD corrections since they are not available beyond this order.

\section{Slepton pair production}
\vspace{-1em}
In this section, we present numerical results for first and second generation left-handed sleptons pair production at the LHC.
A simple rescaling of the cross sections can be applied for the cases of right-handed and/or third generation sleptons, accordingly to their couplings and mixing.
The current experimental exclusion limits bound from ATLAS and CMS experiments, set the masses of first and second generation left-handed sleptons above 550 GeV and 400 GeV respectively, as their analysed data set correspond to integrated luminosities of 139 fb$^{-1}$ and 35.9 fb$^{-1}$ respectively~\cite{Aad:2019vnb,Sirunyan:2018nwe}.

\begin{figure}[h]
\centering
\includegraphics[width=0.32\textwidth]{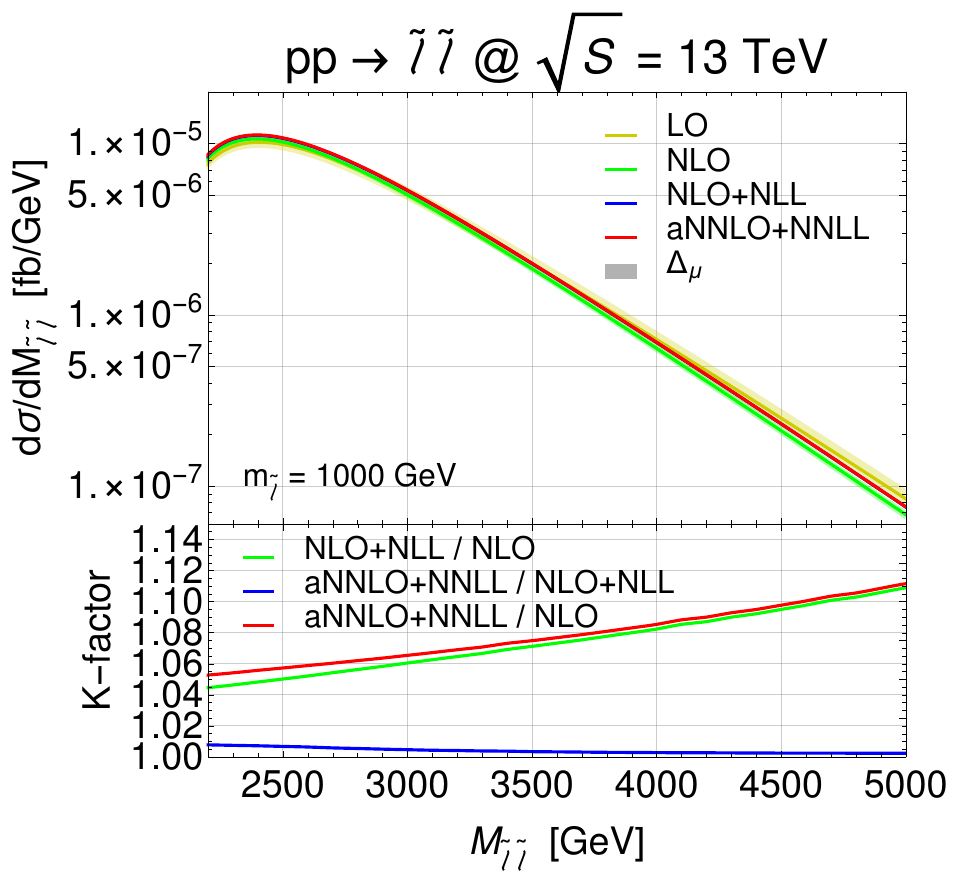}{(a)}
\includegraphics[width=0.43\textwidth]{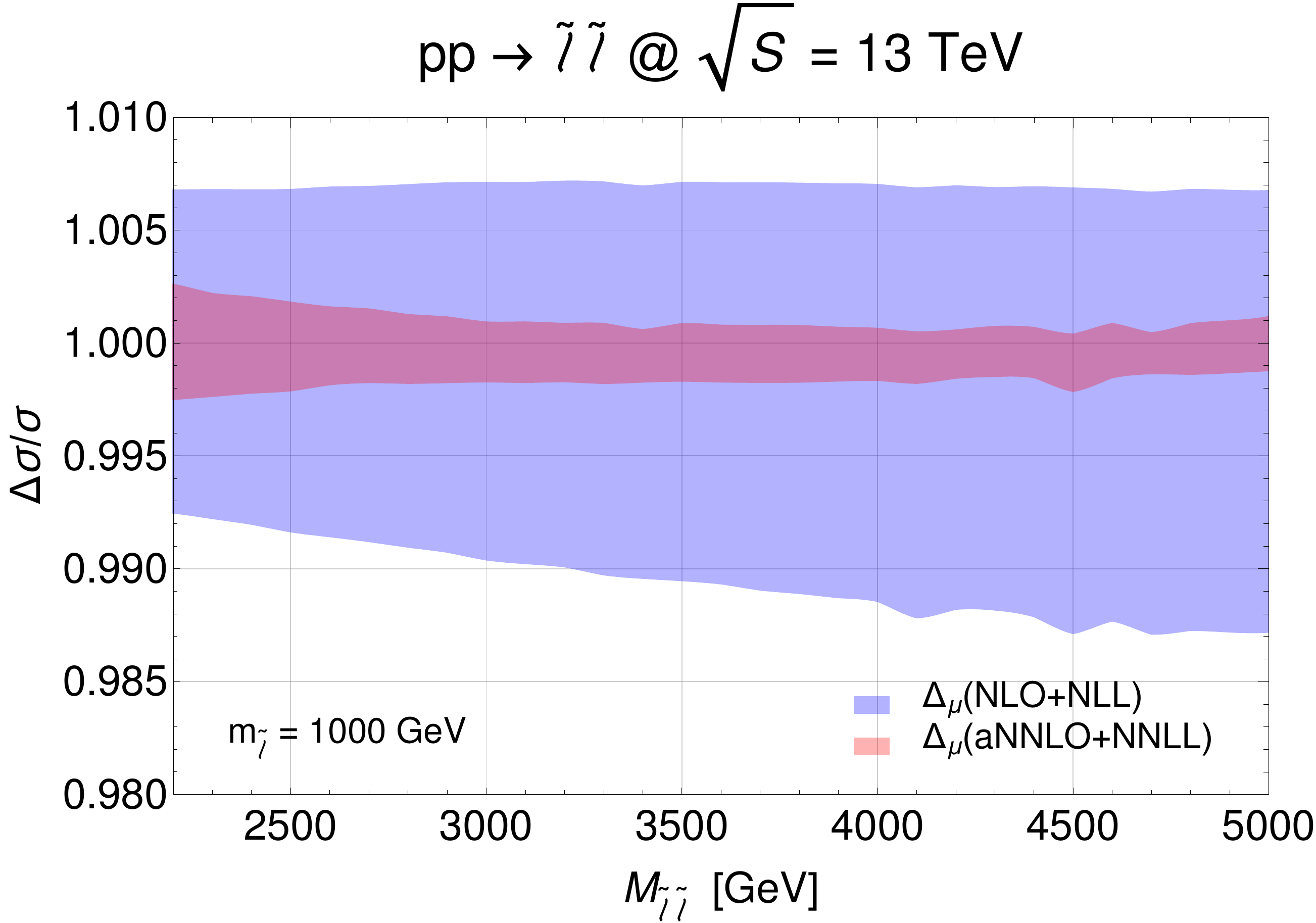}{(b)}
\caption{(a) Invariant-mass distribution for left-handed selectron (or smuon) pair production at the LHC for a fixed slepton mass of 1 TeV and $K$ factors, and (b) its scale uncertainty at NLO+NLL and aNNLO+NNLL.}
\label{fig:Sleptons_Minv}
\end{figure}

In Fig.~\ref{fig:Sleptons_Minv} we show the invariant mass distribution and its scale uncertainty for a fixed slepton mass of 1 TeV.
On the left, we show the results at different pertubative fixed orders and the resummed results together with their $K$ factors.
On the right we can appreciate the considerable reduction of scale uncertainty from mostly below 1\% at NLO+NLL to about 0.1\% at aNNLO+NNLL.

\begin{figure}[h]
\centering
\includegraphics[width=0.32\textwidth]{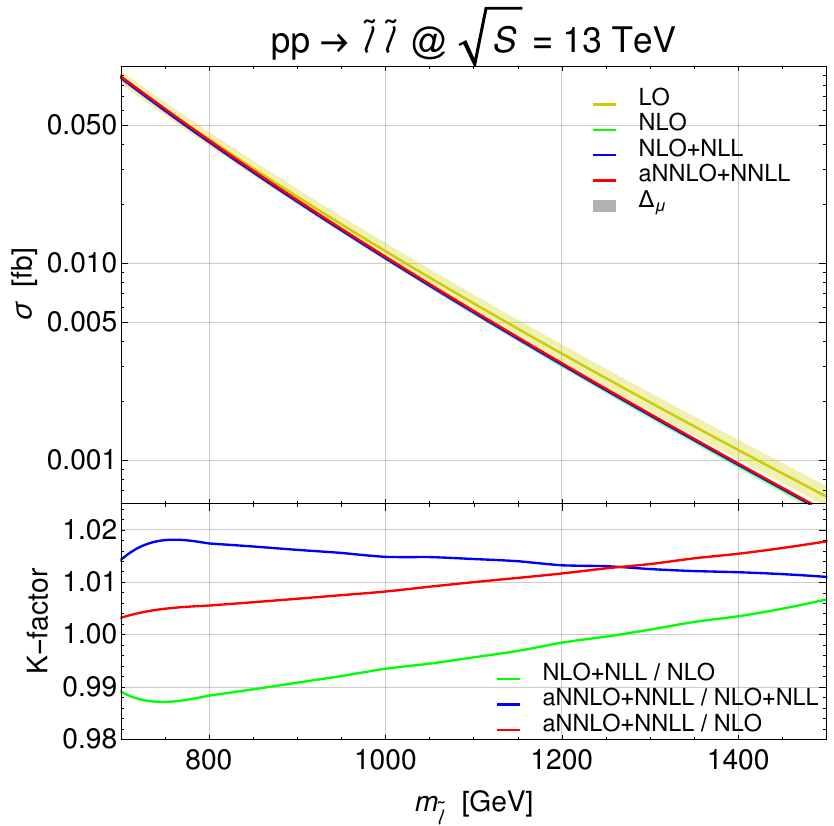}{(a)}
\includegraphics[width=0.45\textwidth]{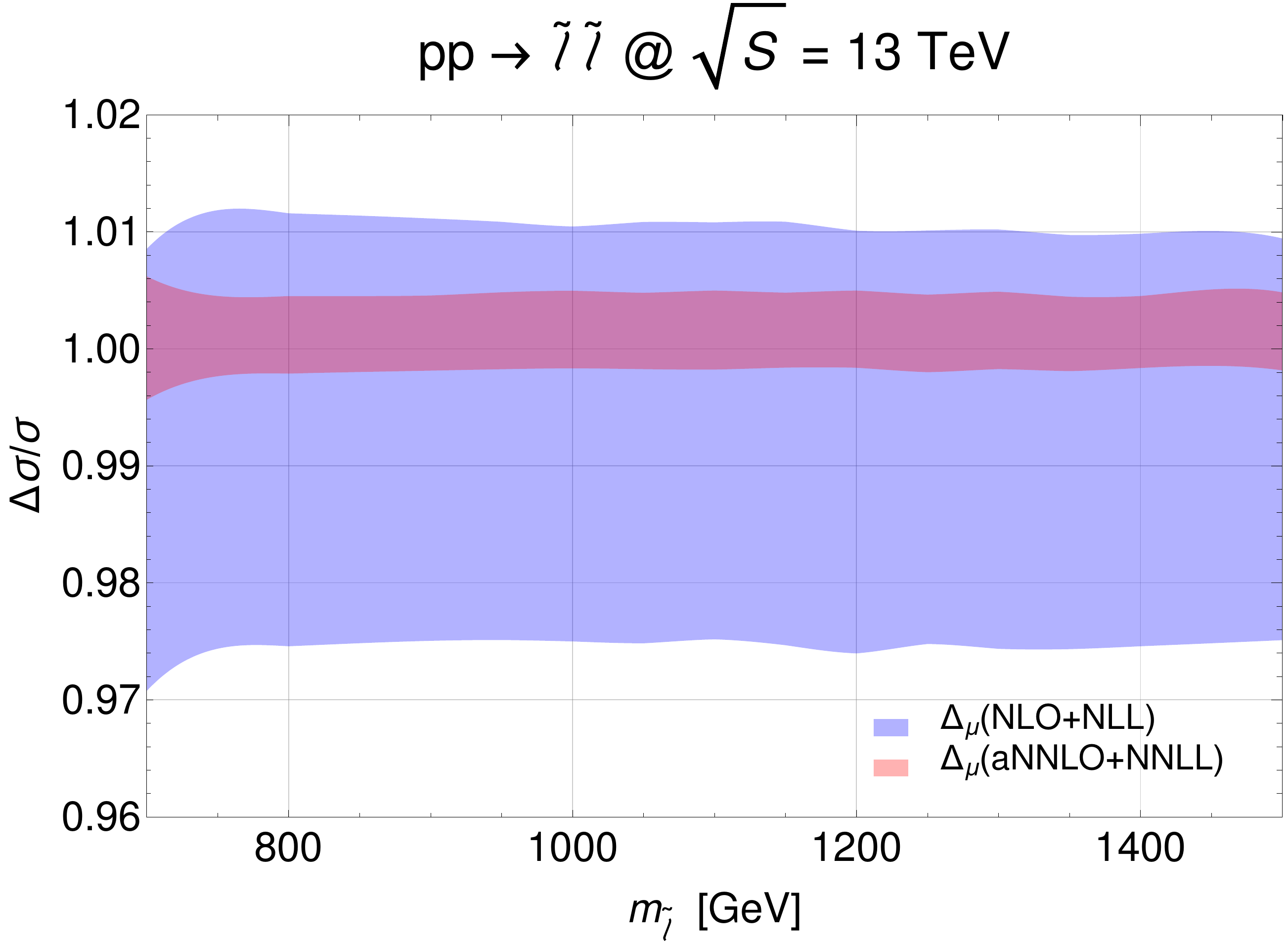}{(b)}
\caption{(a) Total cross section for left-handed selectron (or smuon) pair production at the LHC as a function of the slepton mass, and $K$ factors, and (b) its scale uncertainty at NLO+NLL and aNNLO+NNLL.}
\label{fig:Sleptons_XS}
\end{figure}

In Fig.~\ref{fig:Sleptons_XS} we show similar plots for the total cross section, as function of the slepton mass.
The selected mass range is of direct interest for current and future searches at the LHC since the cross section falls from 0.1 fb to below 1 ab, corresponding to more than 10 events at 700 GeV with an integrated luminosity of 139 fb$^{-1}$, to 3 events at 1 TeV with the LHC Run 3 goal of 300 fb$^{-1}$ and to a few events at 1.5 TeV with the high-luminosity (HL) LHC goal of 3 ab$^{-1}$.
In the plot on the right we observe the reduction of scale uncertainty which uncertainty stretches between -3\% and +1\% at NLO+NLL and between -0.2\% and +0.4\% at aNNLO+NNLL.

\section{Neutralino-chargino production}
\vspace{-1em}
The fermionic superpartners of SM Higgs and gauge bosons can assume different configurations depending on the MSSM parameters.
Following the indications in~\cite{Fuks:2017rio} we generated consistent mass spectra where the neutralinos and charginos are either mostly higgsino-like or gaugino-like.
Here we consider both scenarios and we predict the cross section for neutralino-chargino production where their masses are chosen consistently with the current experimental limits.

\subsection{Mostly Higgsino neutralinos}
\vspace{-0.8em}
Higgsinos masses are required to be small by naturalness arguments.
Furthermore in this configuration, the spectrum features almost degeneracy in the masses of the LSP $\tilde{\chi}_1^0$, the lightest chargino $\tilde{\chi}_1^\pm$ and next-to-lightest neutralino $\tilde{\chi}_2^0$.
Current LHC exclusion limits for almost degenerate higgsino-like pairs $\tilde{\chi}_1^\pm \tilde{\chi}_2^0$ from ATLAS (CMS) collaborations obtained from the analysis of data set with 139 (36) fb$^{-1}$ bound their masses be above 193 (168) GeV~\cite{Aad:2019qnd,Sirunyan:2018iwl}.
Those limits are however strongly dependent on the SUSY model and on the assumed mass splitting. 

\begin{figure}[h]
\centering
\includegraphics[width=0.32\textwidth]{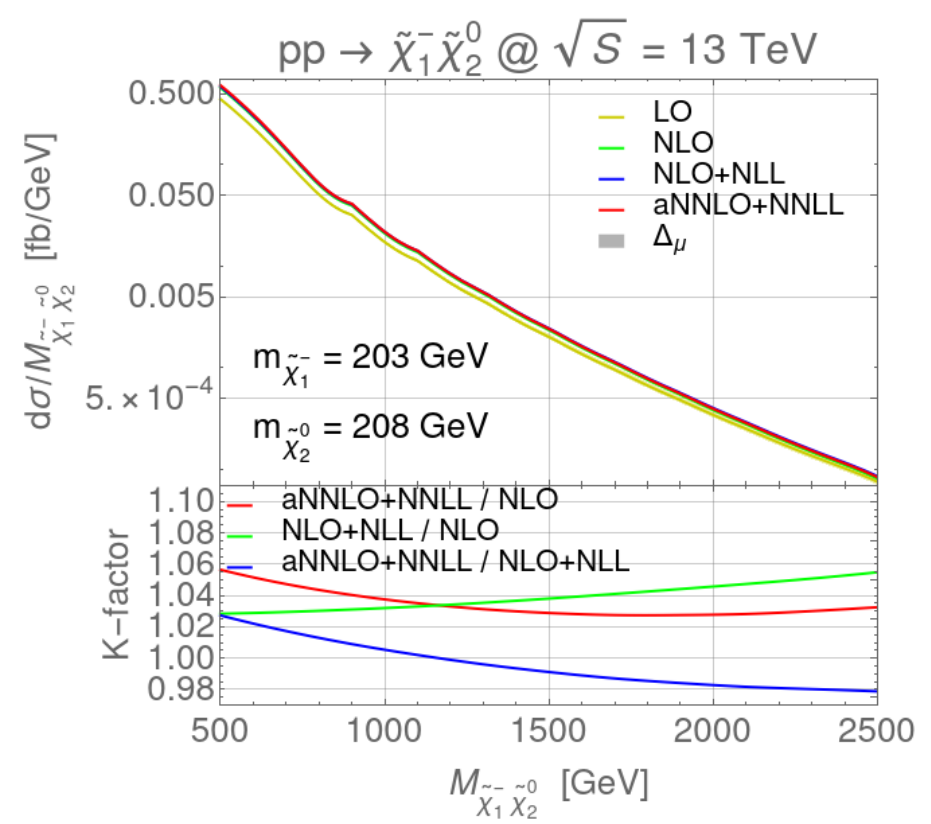}{(a)}
\includegraphics[width=0.40\textwidth]{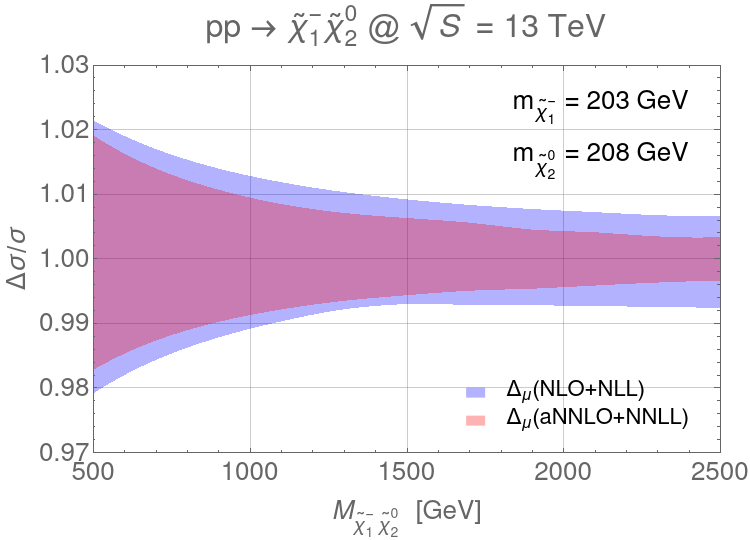}{(b)}
\caption{(a) Invariant-mass distribution for the associated production of higgsino-like charginos and neutralinos with masses of 203 GeV and 208 GeV at the LHC and $K$ factors, and (b) its scale uncertainty at NLO+NLL and aNNLO+NNLL.}
\label{fig:Higgsinos_Minv}
\end{figure}

In Fig.~\ref{fig:Higgsinos_Minv} the masses of neutralino (chargino) are fixed at 208 (203) GeV. On the left is plotted the invariant mass distribution at different perturbative fixed orders and including resummation at NLL and NNLL together with their $K$ factors, while on the right it is shown the scale uncertainty at NLO+NLL and aNNLO+NNLL.
The error bands shrink from $\pm$2.1\% to $\pm$1.8\% and from $\pm$0.6\% to $\pm$0.4\% at low and high invariant masses respectively.

\begin{figure}[h]
\centering
\includegraphics[width=0.32\textwidth]{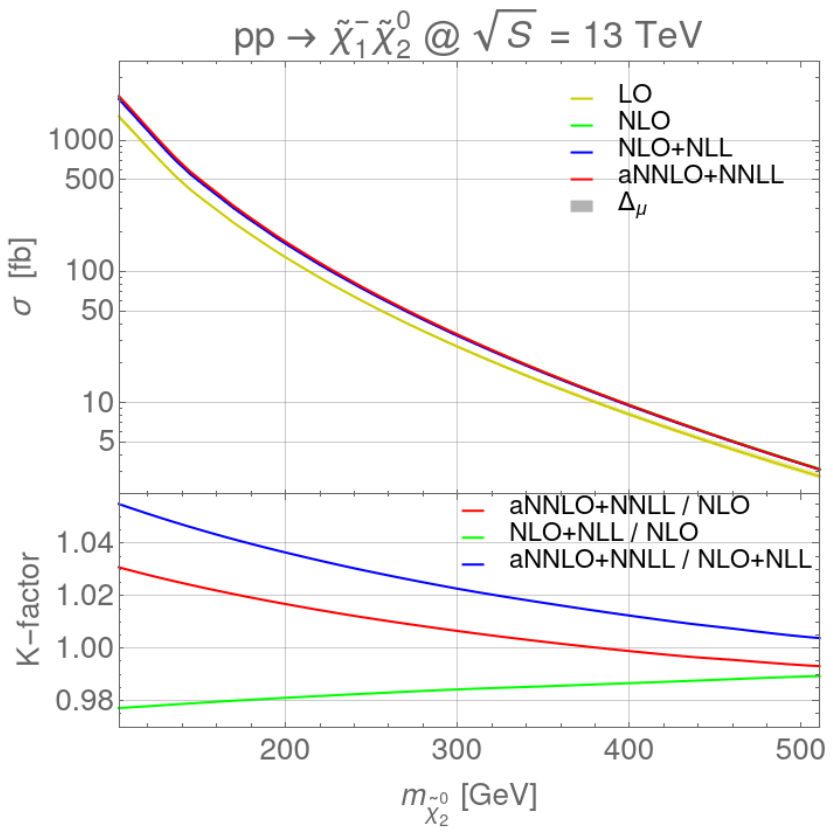}{(a)}
\includegraphics[width=0.45\textwidth]{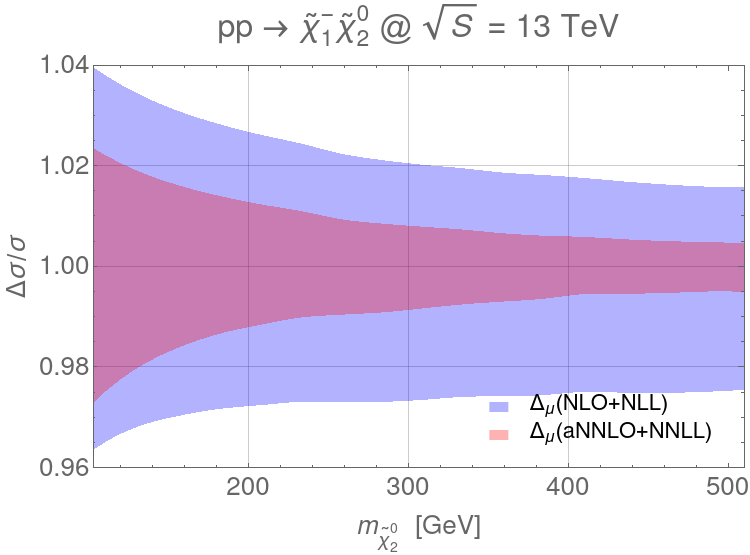}{(b)}
\caption{(a) Total cross section for the associated production of higgsino-like charginos and neutralinos at the LHC as a function of the $\tilde{\chi}_2^0$ mass and $K$ factors, and (b) its scale uncertainty at NLO+NLL and aNNLO+NNLL.}
\label{fig:Higgsinos_XS}
\end{figure}

Similar curves are displayed in Fig.~\ref{fig:Higgsinos_XS}, this time for the total cross section as function of the $\tilde{\chi}_2^0$ mass, ranging between the LEP limit of 103.5 GeV~\cite{Heister:2002mn,Abdallah:2003xe} up to 500 GeV.
Very similar results can be obtained for the final state with a positive chargino $\tilde{\chi}_1^+ \tilde{\chi}_2^0$ and with a pair of charginos $\tilde{\chi}_1^+ \tilde{\chi}_1^-$, with the absolute size of the cross section being larger for a neutral final state, and even more large for a positively charged final state, being the LHC a $pp$ collider.
The plot on the right shows that also in this case there is a visible reduction of the scale uncertainty from about $\pm$2\% at NLO+NLL to $\pm$0.5\% at aNNLO+NNLL for heavy higgsinos.

\subsection{Mostly gaugino neutralinos}
\vspace{-0.8em}
In their mostly gaugino configuration, the next-to-lightest neutralino $\tilde{\chi}_2^0$ and charginos $\tilde{\chi}_1^\pm$ have a large wino component and they are almost degenerate, while the LSP $\tilde{\chi}_1^0$ is mostly bino-like lighter.
The final state $\tilde{\chi}_1^\pm \tilde{\chi}_2^0$ has a large cross section, and the experimental exclusion limits from ATLAS (CMS) obtained with the analysis of data set with integrated luminosity of 36 fb$^{-1}$ exclude chargino masses above 1100 (800) GeV~\cite{Aaboud:2018jiw,Sirunyan:2018lul}.

\begin{figure}[h]
\centering
\includegraphics[width=0.32\textwidth]{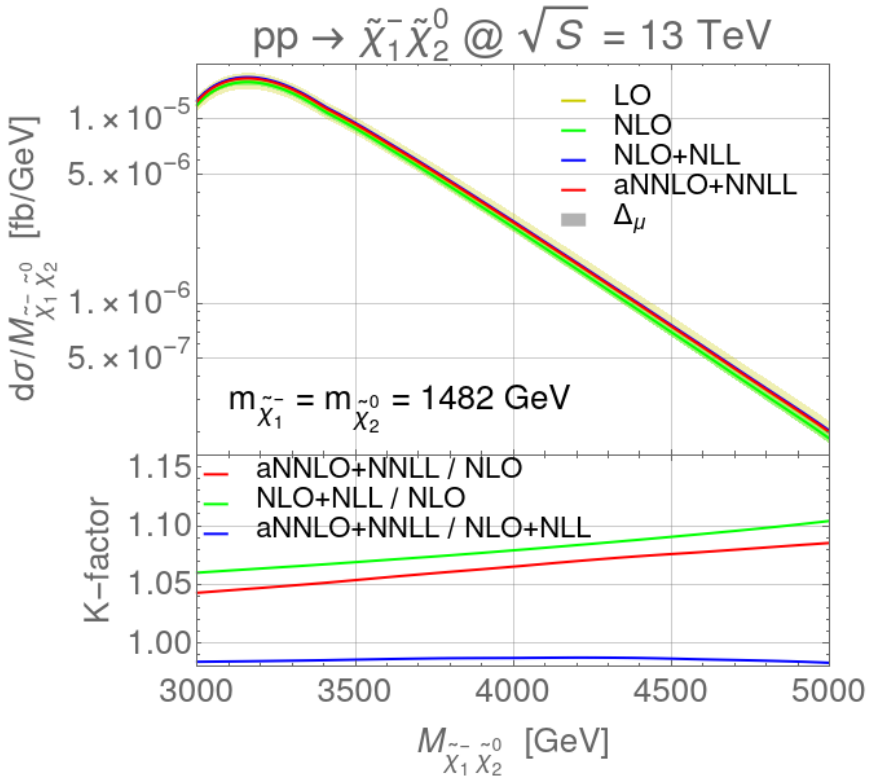}{(a)}
\includegraphics[width=0.40\textwidth]{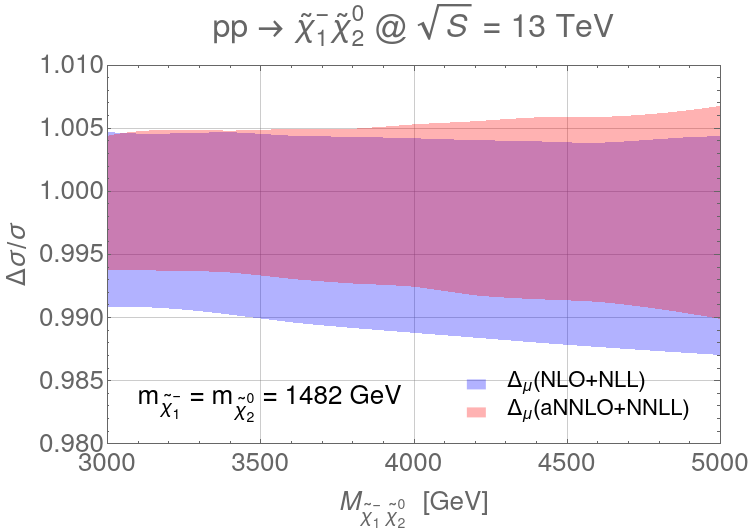}{(b)}
\caption{(a) Invariant-mass distribution for the associated production of gaugino-like charginos and neutralinos with masses of 1482 GeV at the LHC and $K$ factors, and (b) its scale uncertainty at NLO+NLL and aNNLO+NNLL.}
\label{fig:Gauginos_Minv}
\end{figure}

Fig.~\ref{fig:Gauginos_Minv} on the left contains the invariant mass distribution and $K$ factors for the $\tilde{\chi}_1^- \tilde{\chi}_2^0$ final state where the mass of the final state particles if fixed at 1482 GeV, while on the right the scale uncertainty is shown.
At low invariant masses the error band is reduced from $\pm$0.7\% at NLO+NLL to $\pm$0.5\% at aNNLO+NNLL.
At higher invariant masses the reduction of the scale uncertainty is smaller due to the increasing importance of $t$- and $u$-channels contributions.

\begin{figure}[h]
\centering
\includegraphics[width=0.32\textwidth]{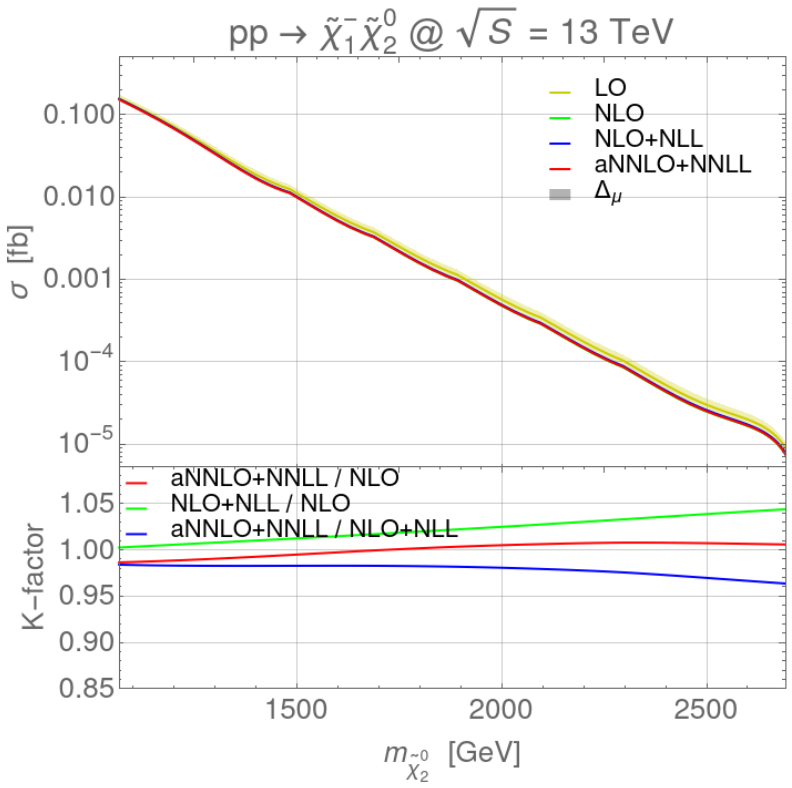}{(a)}
\includegraphics[width=0.43\textwidth]{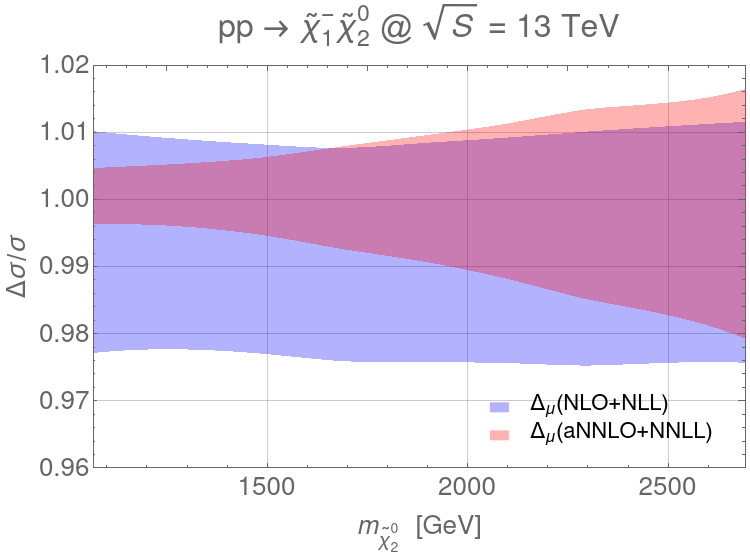}{(b)}
\caption{(a) Total cross section for the associated production of gauginos-like charginos and neutralinos at the LHC as a function of the $\tilde{\chi}_2^0$ mass and $K$ factors, and (b) its scale uncertainty at NLO+NLL and aNNLO+NNLL.}
\label{fig:Gauginos_XS}
\end{figure}

Fig.~\ref{fig:Gauginos_XS} on the left displays the total cross section as function of the $\tilde{\chi}_2^0$ mass and the respective $K$ factors for the same final state.
Similarly as in the higgsino configuration scenario, a larger cross section is obtained when considering the neutral final state $\tilde{\chi}_1^+ \tilde{\chi}_1^-$, and even larger when instead considering the positively charged final state $\tilde{\chi}_1^+ \tilde{\chi}_2^0$.
The $K$ factor curves indicate a better convergence of the perturbative series for lighter gauginos rather than for heavy ones.
This is also reflected in the right plot, where the width of the scale uncertainty error band is about 3\% at NLO+NLL and only 1\% at aNNLO+NNLL, and about 3\% in both cases for light and heavy gauginos respectively.

\section{Conclusions}
\vspace{-1em}
We have presented updated results for the electroweak production of SUSY particles as sleptons and neutralino-chargino pairs, the latter both in their mostly higgsino-like and mostly gaugino-like configurations.
The results have been updated to the current LHC c.o.m. energy of 13 TeV and their theoretical precision has been improved from NLO+NLL to aNNLO+NNLL.
While the impact on the cross sections of the additionally included terms is moderate, we observe a sensible reduction of the scale uncertainties of the predictions that are now around or below the permil level.
The updated calculations have been implemented in the public code RESUMMINO and will be adopted in current and future experimental analysis for SUSY searches at the LHC.

\section*{Acknowledgements}
\vspace{-1em}
\noindent
This work has been supported by the BMBF under contract 05H18PMCC1 and the DFG through the Research Training Network 2149 ``Strong and weak interactions - from hadrons to dark matter''.

\end{document}